\documentclass[aps,prl,twocolumn,superscriptaddress,fleqn,a4paper]{revtex4-1}
\usepackage[ngerman,english]{babel}
\usepackage[utf8]{inputenc}   
\usepackage{xspace}
\usepackage{amsmath,tensor}
\usepackage{array,color}
\usepackage{relsize,exscale}  
\usepackage{textcase}	
\usepackage{soul,color}		
\usepackage{calc} 
\usepackage{makeidx,showidx}
\usepackage{graphicx}
\usepackage{wrapfig}
\usepackage{float}
\usepackage{textcomp}
\usepackage{siunitx}
\usepackage[version=3]{mhchem}
\usepackage{enumitem}
\usepackage{fancyref}
\usepackage[hidelinks=true]{hyperref}
\sethlcolor{green}
\setulcolor{red}
\usepackage{lineno}			

\newcommand{\lsco} {{La$_{2-x}$Sr$_x$CuO$_4$}\@\xspace}

\newcommand{\ybco} {$\ce{YBa2Cu3O_{6+y}}$\@\xspace}

\newcommand{\ybcoE} {$\ce{YBa2Cu4O8}$\@\xspace}

\newcommand{\tc} {\ensuremath{T_{\rm c}}\@\xspace}
\newcommand{\cperp}{\ensuremath{c \bot B_0}\@\xspace}
\newcommand{\cpara}{\ensuremath{{c\parallel\xspace B_0}}\@\xspace}

\newcounter{exex}[section]

\makeatletter\newcommand\listofexamples{\section*{List of Examples}\@starttoc{xmp}}
\newcommand\l@example[2]{\par\noindent#1~\textit{#2}\par}
\makeatother






\makeatletter
\renewcommand\subsection{\@startsection 
	{subsection}{3}{0mm}
	{-\baselineskip}
	{0.5\baselineskip}
	{\centering \textbf }}
\makeatother

\makeatletter
\renewcommand\subsubsection{\@startsection 
	{subsubsection}{3}{0mm}
	{-\baselineskip}
	{0.5\baselineskip}
	{\centering  }}
\makeatother

\makeindex

\begin{document}

\title{Temperature independent pseudogap from $^{17}$O and $^{89}$Y NMR and the single component picture }




\date{\today}
\author{Marija Avramovska }
		\author{Jakob Nachtigal}
		\author{J\"urgen Haase}


\affiliation{University of Leipzig, Felix Bloch Institute for Solid State Physics, Linn\'estr. 5, 04103 Leipzig, Germany}


\begin{abstract}
Nuclear Magnetic Resonance (NMR) is a powerful local quantum probe of the electronic structure of materials, but in the absence of reliable theory the interpretation of the NMR data can be challenging. This is true in particular for the cuprate high-temperature superconductors. Over the years, a large base of NMR data became available, which makes a review of the early interpretation possible. Recently, it was shown that all planar $^{17}$O NMR shift and relaxation data available in the literature point to a temperature independent but doping dependent pseudogap, very similar to what was proposed from the electronic entropy. Here we analyze the anisotropy of the shift and relaxation of planar O to establish whether a single electronic spin component is applicable, since the planar Cu shift anisotropy clearly fails such a description. We find that the orbital shift terms deduced from the data are in agreement with first principle calculations, and the shift data show a temperature independent anisotropy also in agreement with hyperfine coefficients predicted by first principles, which also account for the relaxation anisotropy. Furthermore, we show that the original $^{89}$Y shift and relaxation data are in agreement with the proposed temperature independent pseudogap. This pseudogap depends on doping, but also on the family of materials, and the density of states outside or in the absence of the gap is universal for the cuprates; this suggests that the entropy should be similar for all cuprates, as well. Further consequences will be discussed.

\end{abstract}

\maketitle
\section{Introduction}\label{sec:one}
\begin{figure}
	\includegraphics[width=0.45\textwidth]{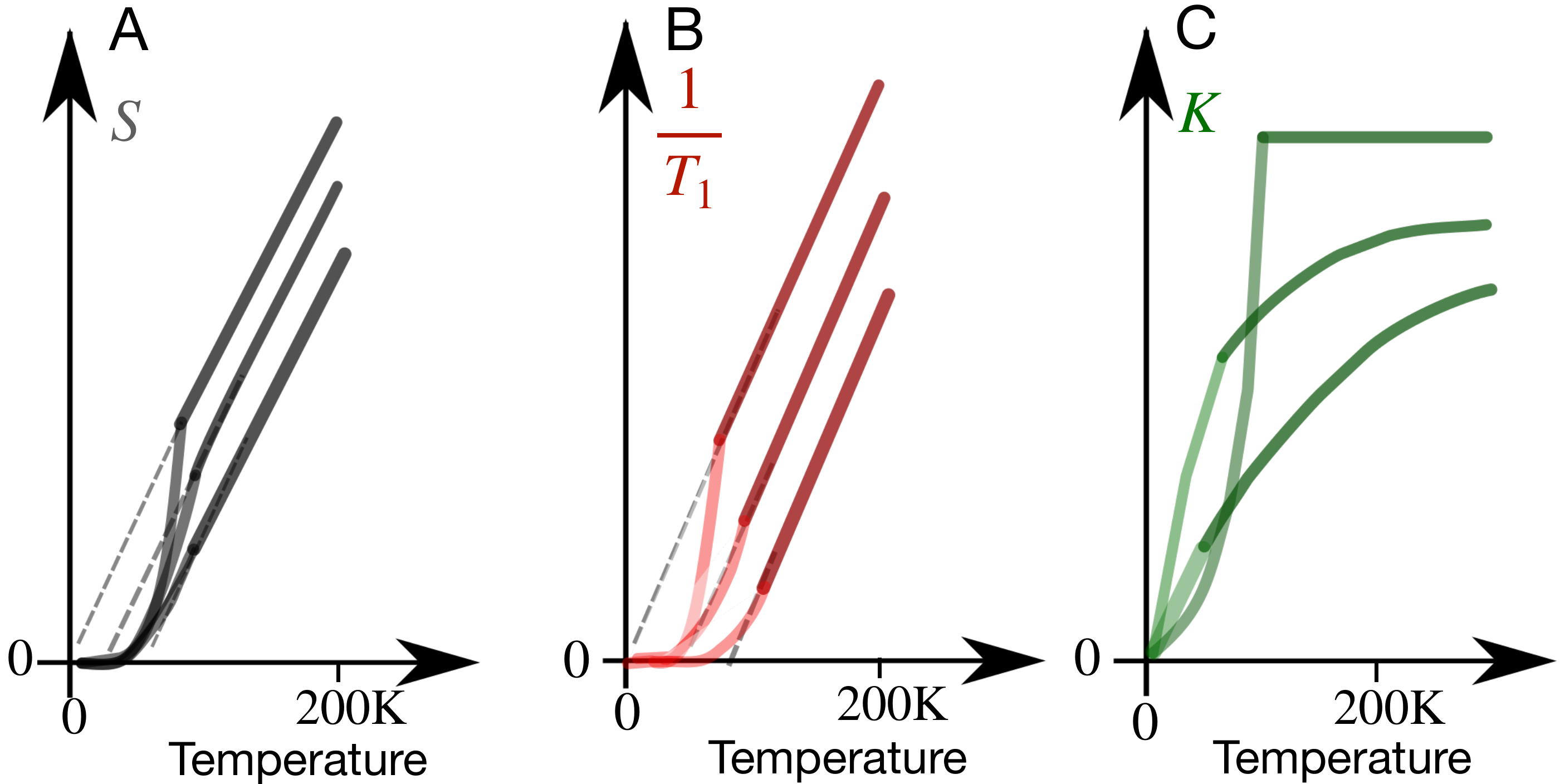}
	\caption{\textbf{A} Sketch of entropy ($S$) vs. temperature ($T$) as adapted from Loram et al. \cite{Loram1998}. \textbf{B} Sketch of nuclear relaxation ($1/T_1$) vs. temperature for different doping levels \cite{Nachtigal2020}; overdoped systems behave metal-like above the critical temperature for superconductivity $T_\mathrm{c}$; at other doping levels the high-temperature slopes remain unchanged. \textbf{C} Sketch of spin shift ($K$) vs. temperature for different doping levels \cite{Nachtigal2020}. Note that the model of the temperature independent pseudogap proposed in \cite{Nachtigal2020} does not consider the superconducting gap.} 
	\label{fig:fig1}
\end{figure}

Nuclear magnetic resonance (NMR), as a bulk probe of material properties with atomic scale resolution, was an early touchstone of understanding cuprate high-temperature superconductors \cite{Bednorz1986}, for reviews see \cite{Slichter2007,Walstedt2008}.  From the  parent antiferromagnets to the overdoped conductors, a number of features were discovered and discussed in terms of classical as well as new theory. For the superconducting materials it was concluded that spin-singlet pairing was behind the loss in spin shift and relaxation below the critical temperature of superconductivity ($T_\mathrm{c}$), as known for classical superconductors \cite{Hebel1957,Yosida1958}. But it was also noticed that the shifts ($K$) and normalized relaxation ($1/T_1T$), start to decrease at temperatures far above \tc, cf.~Fig.~\ref{fig:fig1}, which marked the discovery of the pseudogap (with $^{89}$Y NMR in \ybco) \cite{Alloul1989}. Note that in metals the spin shift ($K$, Knight shift from the Pauli spin susceptibility above \tc), as well as the normalized relaxation ($1/T_1T$) are temperature independent. This metallic behavior is found only for the strongly doped cuprates. The decrease at temperatures above \tc seen in the lower doped materials was attributed to the opening of a spin gap at that temperature. However, the characteristics of the pseudogap did not become clearer as the experimental basis grew over the years, in particular, the thus derived pseudogap temperature is lower that what was found with other probes \cite{Timusk1999}.

Very recently, a different definition of the NMR pseudogap was proposed \cite{Nachtigal2020}, which is the same as the one based on the specific heat data of Loram et al.\@\cite{Loram1998,Tallon2020}. Based on all available planar O data in the literature (more than 35 independent sets of relaxation, and more than 45 sets of shifts) it was found that the high-temperature slopes of the nuclear relaxation with respect to temperature ($\Delta(1/T_1)/\Delta(T)$, cf.~Fig.~\ref{fig:fig1}) are independent on doping, in accordance with a temperature independent pseudogap that opens at the Fermi surface and causes loss of low-energy states. With other words, the states that become available as the Fermi function opens with temperature are metal like and even appear to be ubiquitous to the cuprates \cite{Nachtigal2020}, as the high-doping slope is the same for all materials and the same as in the absence of the gap. Another simple consequence of such behavior is that the NMR shifts, from excitations across this gap, acquire a temperature dependence due to the Fermi function, with the hallmark result that even at the highest temperatures, the gapped, low-energy states are missing. Thus, there is a high-temperature shift offset in the spin susceptibility that depends on doping \cite{Nachtigal2020}. 
The consequences of such a temperature independent pseudogap are sketched in Fig.~\ref{fig:fig1}. 
The lower the doping the more states disappear near the Fermi surface as the size of the pseudogap increases. 

Whether the average doping ($x$, as in \lsco), is the best parameter to describe the pseudogap is not clear. NMR can also measure the local charges at planar Cu ($n_\mathrm{Cu}$) and planar O ($n_\mathrm{O}$) and determine the sharing of the inherent hole as well as the holes added by doping \cite{Haase2004,Jurkutat2014}, and a simple relation was found: $1+ \zeta = n_\mathrm{Cu} + 2n_\mathrm{O}$. Here, $\zeta$ is the doping measured with NMR and follows from $n_\mathrm{Cu}$ and $n_\mathrm{O}$. $\zeta$ is found to be similar to $x$ for \lsco. This is expected since the equation describes the stoichiometry of the materials. However, there are slight differences between $\zeta$ and what was concluded for other materials based on estimates of doping. Interestingly, the sharing  of the planar hole content, i.e. $n_\mathrm{Cu}/n_\mathrm{O}$, varies strongly among the cuprate families and appears to set various material properties \cite{Jurkutat2019b}. Most notably, the maximum \tc of the hole doped cuprates is proportional to $n_\mathrm{O}$. This dependence of cuprate properties on the hole sharing was very recently proven by theory (cellular DMFT) \cite{Kowalski2021}, lending strong support to the NMR findings and the importance of $\zeta, n_\mathrm{O}, n_\mathrm{Cu}$. These parameters may be of importance for the understanding of shift and relaxation, as well. While it is not new that doping plays an important role, there are also family dependences that track those of sharing the charges, such as the planar Cu shift anisotropy. This behavior is at the root of the fact that these shifts cannot be explained with a single temperature dependent spin component \cite{Haase2017,Avramovska2019}. A two component behavior was also proposed for O since the planar O shifts have a very different temperature dependence than that of the shift distributions \cite{Pavicevic2020}. 

It is therefore of great importance to investigate the planar O shift and relaxation anisotropies, which could not be accomplished with the first account of the temperature independent pseudogap \cite{Nachtigal2020}. While the number of datasets we could allocate for addressing this question is more limited compared to what was measured with the magnetic field along the crystal $c$-axis \cite{Nachtigal2020}, the data are representative for the cuprates. As we will show, the planar O shift and relaxation anisotropies can be understood fully within a single spin component scenario, with the orbital shifts in agreement with what was predicted by first principle calculations \cite{Renold2003}. Even the anisotropies follow from an anisotropic hyperfine constant that is in agreement with what was calculated by first principles \cite{Huesser2000}. Finally, we will show that the original $^{89}$Y NMR data \cite{Alloul1989} are in agreement with a temperature independent pseudogap, and we discuss consequences.

\section{Data collection}\label{sec:two}
We collected data for planar oxygen with all 3 directions measured in an intensive literature search and found the materials listed in the Appendix. The planar O shift tensor's main principle axis is along the Cu-O-Cu $\sigma$-bond (denoted by $K_{\parallel\sigma}$). The other two axes are assumed to be perpendicular to this bond, and along the crystal $c$-axis ($^{17}\hat{K}_{\perp\mathrm{c}}$) and $a$-axis ($^{17}\hat{K}_{\perp\mathrm{a}}$). The relaxation was also measured with the field along these axes. The temperature dependences of shift and relaxation were manually extracted from the figures, including error bars. For the sets of data that do not share the exact same temperature points for all three directions of the field, the original data were interpolated using a linear fit between the two respective data points. No explicit error bars were reported for Tl$_2$Ba$_2$CuO$_y$ \cite{Kambe1993} and 
YBa$_{1.92}$Sr$_{0.08}$Cu$_3$O$_7$ \cite{Horvatic1989}, and we assume they are similar to those of the other cuprates. A source of systematic uncertainty across all materials is the dependence on sample preparation procedures, in particular the fact that the doping level may not be known with high precision and oxygen isotope exchange may not be homogeneous for single crystals or even affect the actual O doping level. For example, materials from different groups can show somewhat different shifts, as is the case for La$_{1.85}$Sr$_{0.15}$CuO$_4$ data from Kitaoka et al. \cite{Kitaoka1988}, Ishida et al. \cite{Ishida1991}, and Singer et al. \cite{Singer2002}. Interestingly, this is also the case even for the stoichiometric compound \ybcoE (see figures 2 and 3 in \cite{Nachtigal2020}). If the shift reference is not stated explicitly we assume it is water (a very convenient choice). As before \cite{Nachtigal2020}, we show the total, uncorrected shifts with respect to this reference and determine the orbital shifts later to keep data analysis transparent. 

\begin{figure*} [ht]
	\includegraphics[width=\textwidth]{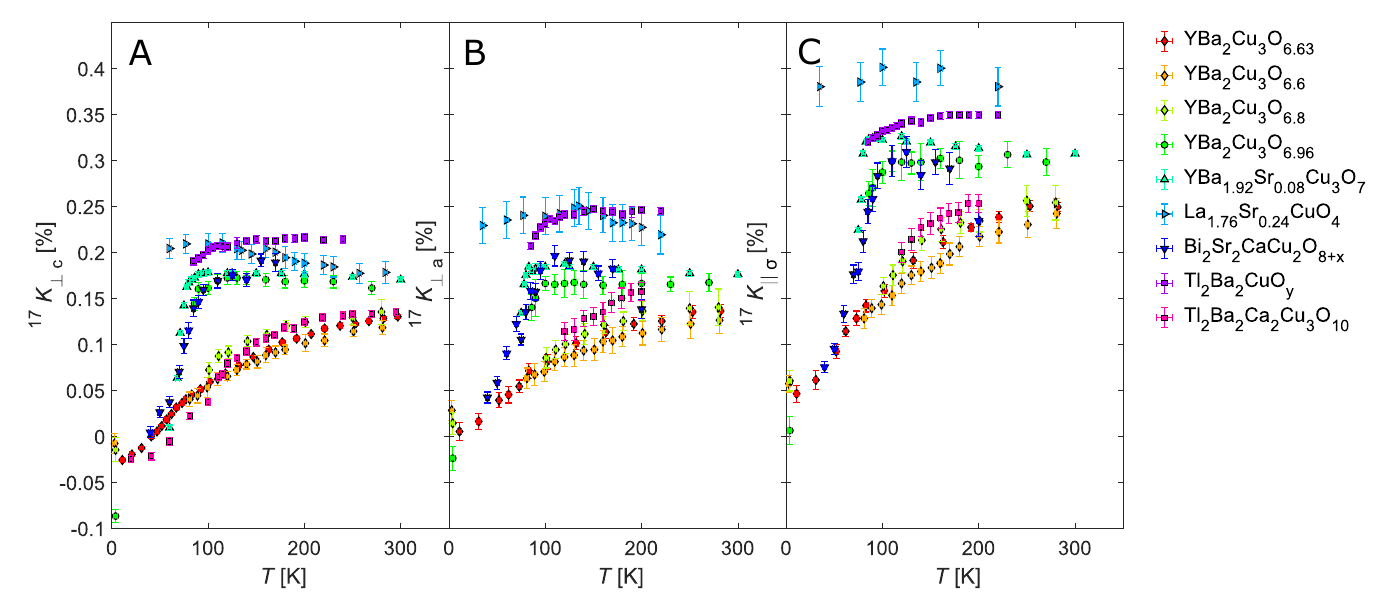}
	\caption{Temperature dependences of the planar oxygen shifts for different cuprate-families for all three directions of the tensor axes. \textbf{A} $^{17}\hat{K}_\mathrm{\perp c}(T)$, \textbf{B} $^{17}\hat{K}_\mathrm{\perp a}(T)$, and \textbf{C} ${^{17}\hat{K}}_{\parallel\sigma}(T)$. Optimally and overdoped systems behave metal-like. Underdoped systems show the expected high-temperature behavior with a doping dependent shift offset. Some materials show a negative spin shift at low temperatures. The highest shift range is observed for $K_{\parallel\sigma}$ with \SI{0.40 \pm 0.05}{\%}. Error bars are taken from the literature. For ${K}_\mathrm{\perp c}(T)$ only subset of the data from \cite{Nachtigal2020} is shown as no data for the other directions are available. For the references, see Appendix.}
	\label{fig:fig2}
\end{figure*}

\section{Planar oxygen shifts and relaxation}\label{sec:three}
\subsection{Shifts}
While we have only a limited number of shifts, the dependences for ${^{17}\hat{K}}_{\perp\mathrm{c}}$ in Fig.~\ref{fig:fig2} are quite representative of the cuprates if one compares to the related much more abundant plots in \cite{Nachtigal2020}. In Fig.~\ref{fig:fig2} one finds high-temperature doping-dependent offsets and an onset of the temperature dependence already far above \tc for the underdoped materials, and the shifts nearly vanish at the lowest temperatures without showing special behavior at the superconducting transition temperature. The shifts for the overdoped materials are nearly temperature independent at high temperatures, and drop rapidly near \tc to similar low temperature values. For all 3 directions of the shift tensor we observe this behavior.

\begin{figure*}
	\includegraphics[width=\textwidth]{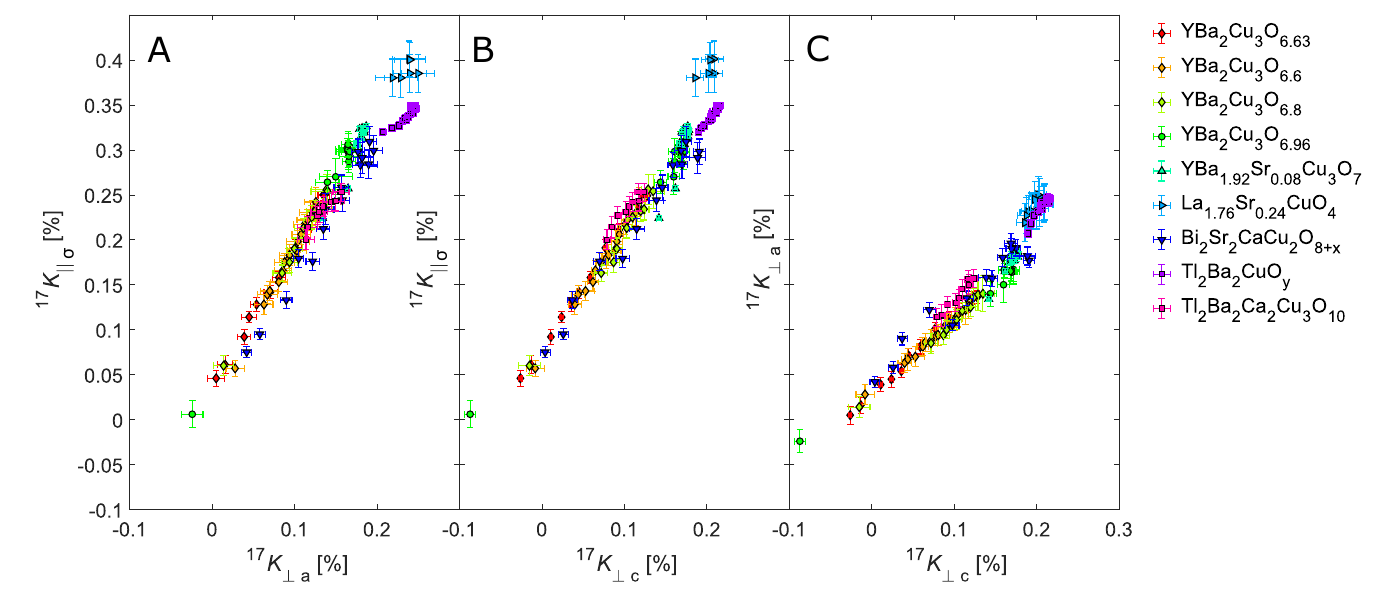}
	\caption{Temperature dependences of the total planar O magnetic shifts from Fig. \ref{fig:fig2}, plotted against each other. \textbf{A} $^{17}\hat{K}_{\parallel \sigma}$ vs. $^{17}\hat{K}_\mathrm{\perp a}$, \textbf{B} $^{17}\hat{K}_{\parallel \sigma}$ vs. $^{17}\hat{K}_\mathrm{\perp c}$ and \textbf{C} $^{17}\hat{K}_\mathrm{\perp a}$ vs. $^{17}\hat{K}_\mathrm{\perp c}$. If one subtracts the orbital shifts, the shifts are proportional to each other in concordance with a single component picture. Note that one may not conclude with this certainty if considering only a single material.}
	\label{fig:fig3}
\end{figure*}

We now use the data from Fig.~\ref{fig:fig2} and plot the various shifts against each other, with the results shown in Fig.~\ref{fig:fig3}. Given the error bars, one concludes on largely temperature independent slopes, i.e. the shift anisotropy appears to be temperature independent. Furthermore, the slope for the two perpendicular shifts is close to 1, while plots involving the shift along the $\sigma$-bond have a larger slope, about 1.45.

The total magnetic shift can be written as,
\begin{equation}\label{eq:shift1}
{^{17}\hat{K}}_\alpha(T) = {^{17}K}_{\mathrm{L}\alpha} + {^{17}K}_{\alpha}(T),
\end{equation}
where ${^{17}K}_{\mathrm{L}\alpha}$ is the orbital shift, and ${^{17}K}_{\alpha}(T)$ denotes the spin shift. 
For a single spin component susceptibility, $\chi(T)$, we have for the spin shift,
\begin{equation}\label{eq:hyper1}
{^{17}K}_\alpha(T) = C_\alpha \cdot \chi(T)\equiv (c_\mathrm{iso} + c_\mathrm{dip,\alpha}) \cdot\chi(T),
\end{equation}
where we assumed the hyperfine coefficient $C_\alpha$ to be the sum of an isotropic term from the O 2s orbital and a (traceless) dipolar term from the 2p orbitals. From the symmetry of the O $2p$ orbitals one expects $C_{\mathrm{dip}\parallel\sigma} \approx 2 |C_{\mathrm{dip}\perp c,a}|$. 
The question arises whether the low temperature offsets in Fig.~\ref{fig:fig3} are due to the orbital shifts. 
Given the uncertainty from a diamagnetic response below \tc, we verified with the high-temperature shifts for those materials that have nearly temperature independent high-temperature shifts, that the offset between the shifts at high temperatures is the same as that found from the low-temperature offsets. This means, a possible diamagnetic contribution below \tc can be neglected. This is in agreement with the fact the the correction from diamagnetism should be smaller than $0.01\%$ \cite{Haase2012}. Then, one would argue with Fig.~\ref{fig:fig3} that the orbital shift is only significant for \cpara were it is appears to be negative. Note that in panel ${\bf A}$ the slope is nearly 1 and the offset is nearly 0, and any diamagnetic contribution should be nearly isotropic in the plane.

Indeed, first principle calculations of the orbital shifts for \lsco \cite{Renold2003} report ${^{17}K}_{\mathrm{Lc}}$ = \SI{-0.034\pm 0.020}{\%}, ${^{17}K}_{\mathrm{La}}$ = \SI{-0.011\pm 0.010}{\%}, and ${^{17}K}_{\mathrm{Lb}}$ = \SI{-0.004\pm 0.005}{\%}. Thus, the straight lines observed in Fig.~\ref{fig:fig3} are expected and we deduce the following slopes: $C^{\mathrm{exp}}_{\perp\mathrm{a}}/C^{\mathrm{exp}}_{\perp \mathrm{c}} \approx {\SI{0.90\pm 0.15}{}}$, $C^{\mathrm{exp}}_{\parallel\sigma}/C^{\mathrm{exp}}_{\perp a} \approx {\SI{1.45\pm 0.10}{}}$. These numbers are in very good agreement with what we find for \lsco and even the other cuprates. Thus this set of orbital shifts applies to all cuprates to a good approximation.

One should note that the linewidths for planar O (that limit the precision of the shifts) typically increase drastically as the temperature is lowered, even much in excess of what is expected from the field variation in the mixed state, making the low $T$ shift values less reliable. Also the penetration depth decreases while the relaxation slows down drastically at low temperatures, making measurement tedious. These are the likely reasons why we find the most low-temperature points for the shifts ${^{17}K}_{\perp\mathrm{c}}$ of \ybco and \ybcoE, since these are also the materials with smallest linewidths. 

%


The magnetic hyperfine coefficients have been determined by first principle calculations, as well \cite{Huesser2000}. It was found that
 $c_\mathrm{iso} = 1.3$, $c_{\mathrm{dip}\parallel\sigma} = 0.37$ (in atomic units), whence there is the factor of 1.5 for $C_{\parallel\sigma}/C_{\perp c,a} $, which is in  agreement with what we find from experiment:  ${^{17}K}_{\parallel\sigma}/{^{17}K}_{\perp\mathrm{a}}\approx  \SI{1.45\pm 0.1}{}$ and  ${^{17}K}_{\perp\mathrm{a}}/{^{17}K}_{\perp\mathrm{c}}\approx \SI{0.90\pm 0.15}{}$. As the plots in panels {\bf A, B} of Fig.~\ref{fig:fig4} show, there is no other significant contribution to the shifts above \tc (below \tc the shifts are small and the error bars for their ratio becomes very large).
 
To summarize, the planar O shifts are fully accounted for by a single spin component coupled to the nucleus with predicted hyperfine coefficients and orbital shifts.  The dependence of the spin susceptibility on doping does not affect the anisotropy of the shifts. This is also true, as expected, for the orbital shifts and the hyperfine coefficients.\par\medskip

\subsection{Relaxation}
\begin{figure}[h!]
	\includegraphics[width=0.5\textwidth]{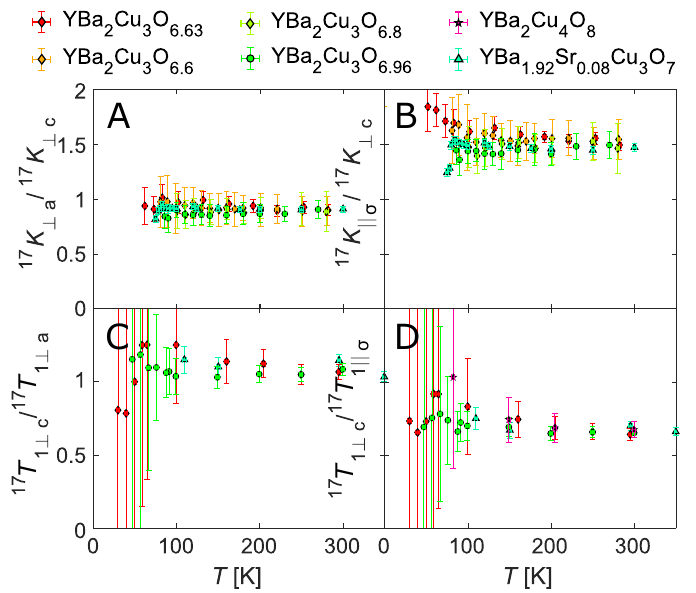}
	\caption{ Temperature dependences of the, \textbf{A, B} shift anisotropies, and \textbf{C, D} of the corresponding relaxation anisotropies. The experimentally obtained shift and relaxation ratios are in close agreement with the predicted hyperfine coefficients (see main text).}
	\label{fig:fig4}
\end{figure}

Accounts of the planar oxygen relaxation anisotropy are sparse. We found mainly data for \ybco \cite{Martindale1993,Horvatic1993,Suter1997} and the data are shown in panels {\bf C, D} of Fig.~\ref{fig:fig4}. Above \tc, the relaxation anisotropy measured perpendicular to the $\sigma$-bond is about 1, and the ratio of the values measured parallel to the $\sigma$-bond and perpendicular to it is about 0.7.

In a simple picture \cite{Pennington1989} the relaxation rates are given by the field fluctuations perpendicular to the direction of the quantization axis (the external field) and we have, for example,
\begin{equation}\label{eq:relaxation}
		1/T_{1\perp\mathrm{c}}=\frac{3}{2}\gamma^2[\left<h_{\parallel\sigma}(t)^2\right>+\left< h_{\perp\mathrm{a}}(t)^2\right>]\tau_0
\end{equation}
where $h_{\parallel\sigma}(t)$ and $h_{\perp\mathrm{a}}(t)$ denote the fluctuating magnetic fields and $\tau_0$ is their correlation time. One can write $\left<h_{\alpha}(t)^2\right>$ in terms of the hyperfine coefficients as $\left<h_{\alpha}(t)^2\right> ={C_{\alpha}^2}/{\gamma_n \hbar} \cdot \left<S^2\right>$, where $\gamma$ is the gyromagnetic ratio of the nucleus and $\left<S^2\right>$ is the expectation value from the electronic spin. With $C^{\mathrm{exp}}_{\perp\mathrm{a}}/ C^{\mathrm{exp}}_{\perp\mathrm{c}} \approx 0.9$ and $C^{\mathrm{exp}}_{\parallel\sigma}/C^{\mathrm{exp}}_{\perp\mathrm{c}} \approx 1.45$ and expressions similar to \eqref{eq:relaxation} for the other directions of measurements, we expect the relaxation perpendicular to the $\sigma$-bond to be isotropic, but $(1/T_{1\parallel\sigma})/(1/T_{1\perp\mathrm{c}}) \approx 0.62$. This is again in good agreement with the observations.

While the error bars below \tc grow, it was shown for \ybcoE below about \SI{200}{K} that quadrupolar relaxation is present, as well \cite{Suter2000}. This additional term is still within error bars of Fig.~\ref{fig:fig4} {\bf C} and {\bf D}.

This leads us to conclude that also the relaxation for the planar O nucleus is largely described by simple on-site fluctuating spin (in the long wavelength limit) with hyperfine coefficients very close to what is predicted by first principle calculations. No special assumptions about filter functions for transferred spin are necessary \cite{Huesser2000}.

\begin{figure}
	\includegraphics[width=0.45\textwidth]{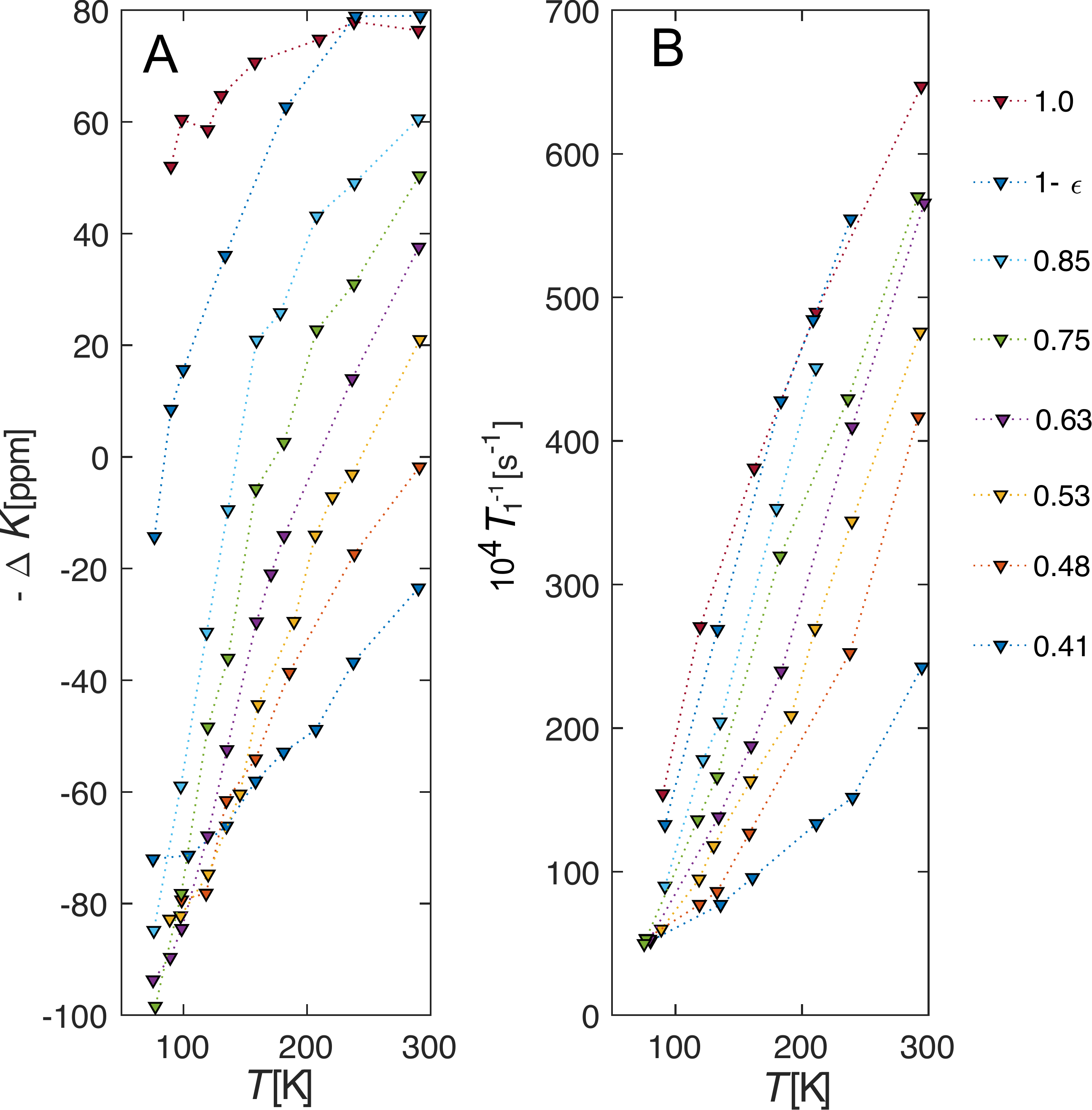}
	\caption{Temperature dependences of the $^{89}$Y shifts, \textbf{A}, and relaxation, \textbf{B}, from Alloul et al. \cite{Alloul1989} are also in support of a temperature independent pseudogap (see main text).}
	\label{fig:fig5}
\end{figure}

\section{Pseudogap and $^{89}$Y NMR of YB\lowercase{a}$_2$C\lowercase{u}$_3$O$_{7-\delta}$}
The $^{89}$Y NMR data that mark the discovery of the pseudogap by Alloul et al. \cite{Alloul1989} were only taken above \tc so that low-temperature values are not available for this nucleus (except for the near optimal doped material \cite{Barrett1990b}), and mostly only available for powders. We reproduce the data in Fig.~\ref{fig:fig5}. Despite the uncertainties, shift and relaxation  show the typical features of a temperature independent pseudogap, as well. Thus the recent conclusions are in agreement with the $^{89}$Y data.

\section{Discussion and Conclusion}

We showed previously \cite{Nachtigal2020} that the oxygen shift and relaxation data, available in abundance as measurements with the field along the crystal $c$-axis, can be modeled by a temperature independent gap at the Fermi surface of a simple metal (using a single spin component). The density of states outside the gap is ubiquitous for the cuprates, this includes the density of states when the gap is fully closed. Since this behavior is similar to measurements of the electronic entropy \cite{Loram1998,Tallon2020} in a few materials, the same doping dependent entropy should govern all cuprates.

We also know from recent Cu NMR analyses that the observations in terms of NMR are more complicated for this nucleus \cite{Haase2017,Avramovska2019}, in particular, the Cu shift anisotropy. Therefore, we focused here on the planar O shift anisotropy that was not analyzed previously. We present literature data where the shifts and relaxation were measured for all three directions of the magnetic field with respect to the Cu-O bond and look at the anisotropies. We find that both, shift and relaxation obey the same single spin component scenario, in agreement with what is expected from the hyperfine coefficients calculated in \cite{Huesser2000}, who report: $C_{\perp a}/C_{\perp c} \approx 0.93$ and $C_{\parallel\sigma}/C_{\perp c,a} \approx 1.5$ while the experimental shifts give $C_{\perp \mathrm{c}}^{\mathrm{exp}}/C_{\perp \mathrm{a}}^{\mathrm{exp}} \approx 0.9$ and $C_{\parallel\sigma}^{\mathrm{exp}}/C_{\perp \mathrm{c,a}}^{\mathrm{exp}} \approx 1.45$. Furthermore, we find the anisotropic orbital shifts to be rather independent on doping and family, and also in agreement with what was predicted from first principles \cite{Renold2003}.

For all oxygen relaxation data \cite{Nachtigal2020}, independent of material, the un-gapped relaxation rate for \cpara has a slope of $\SI{0.35}{/Ks}$ with temperature (this is also the rate determined from relaxation outside the gap for all doping levels). In case of a simple Fermi liquid the {Korringa} relation holds \cite{Korringa1950}: $T_1TK_s^2=\kappa ({\gamma_e}/{\gamma_n})^2 {\hbar}/{4 \pi k_B}$. The prefactor $\kappa$ is usually introduced to describe electronic correlations.
For the  $^{17}$O shifts, where we know both, the orbital shifts and the hyperfine coefficients, we obtain that the Korringa law holds for the metallic density of states with $\kappa_\mathrm{O} \approx 1.8$, note that we are comparing, e.g., ${^{17}K}_{\perp\mathrm{c}}^2 \propto C_{\perp\mathrm{c}}^2 $ with $1/{^{17}T}_{1\parallel \sigma} \propto \left[C_{\perp\mathrm{c}}^2 + C_{\perp\mathrm{a}}^2 \right]$. 
In \cite{Alloul1989}, it was found that the $^{89}$Y data have $\kappa_\mathrm{Y} = 5.3$. Remember that for the $^{89}$Y shifts the orbital contributions are not precisely known. 

It was also found that for some highly overdoped cuparates the Korringa law can explain the relation between Cu shift and relaxation \cite{Avramovska2019}. This was taken as an indication that, differently from what was believed earlier, the relaxation is \emph{not} enhanced over what one can expect from the shifts, as the doping and material independent normalized relaxation rate  of $1/{^{63}T}_{1\perp}T \approx \SI{21}{/Ks}$ already indicates \cite{Jurkutat2019}. Rather, it was shown \cite{Avramovska2019} that the shifts must be suppressed in the underdoped materials. An explanation for the Cu shift suppression could be that the Cu shifts are affected by the pseudogap, much like the O shifts, however the Cu relaxation rates are not.
Moreover, the universal Cu relaxation rate might be connected to the ubiquitous density of states seen with planar O NMR that leads to $1/{^{17}T}_{1\mathrm{c}}T\approx\SI{0.35}{/Ks}$. This normalized rate is proportional to $\left[C_{\perp\mathrm{a}}^2 + C_{\parallel\sigma}^2\right]$ and if we normalize it to $C_{\perp\mathrm{c}}$ we have \SI{0.21}{/Ks}. With a ratio of ${^{63}\gamma}/{^{17}\gamma} \approx 1.96$ we conclude that an effective hyperfine coefficient for Cu should be about $5.1\cdot C_{\perp\mathrm{c}}$, or a spin shift of about $1.2\%$, which is significantly larger than the high-temperature value of $^{63}K_\perp \sim 0.7\%$. Note that the Cu shifts have a large anisotropy that is family dependent. Only for \cperp the orbital shift appears to be reliable \cite{Avramovska2019}. Furthermore, since $1/^{63}T_{1\perp}T \approx \SI{21}{/Ks}$ (caused by the sum of in-plane and out-of-plane fluctuating field components) is rather material independent in contrast to $1/^{63}T_{1\parallel}$ that varies among the materials, the out-of-plane (\cpara) hyperfine coefficient should dominate the Cu relaxation, which is related to $^{63}K_\parallel$. A two-component model that accounts for the relaxation was proposed recently \cite{Avramovska2020}. Nevertheless, work is underway that addresses  Cu shifts and relaxation and comparison with the new results.

The size of the pseudgap was discussed in figure 9 of \cite{Nachtigal2020} and it showed a complicated family dependence. Whereas \ybco has no sizable pseudogap at optimal doping, other cuprates do show a pseudogap. Interestingly, optimally doped \ybco has the largest $\zeta$ \cite{Jurkutat2021}, meaning that it already on the overdoped side in terms of the NMR doping $\zeta$. 
Since the precise meaning of this material dependency ($n_\mathrm{Cu}, n_\mathrm{O}, \zeta$) is not fully understood, it is difficult to draw further conclusions on relations with the pseudogap or \tc. 

Earlier, a two component picture for the O shift was invoked by comparing the different temperature dependences of the shifts and linewidths \cite{Pavicevic2020}. The latter are related to the spatial charge variation in the CuO$_2$ plane. Now, with the knowledge that a variation of doping must lead to a variation of the pseudogap, an inhomogeneous pseudogap easily explains the differences between shifts and linewidths. It is the susceptibility that depends on two parameters (temperature and doping). 

Different temperature dependences of the O axial and isotropic shifts were reported for \ybcoE by Machi et al. \cite{Machi2000}. This  finding seems to contradict a single susceptibility scenario, as well. We observe,  similarly as in \cite{Machi2000} that the oxygen axial shifts, for all the available data, are largely temperature independent, whereas the isotropic shift (or ${^{17}K}_\mathrm{c}$) has a substantial temperature dependence already above \tc. However, we find the overall uncertainty in the axial shift data too large to confirm this scenario.
 
To conclude, from the study of the anisotropies of planar O relaxation and shift we find that the data can be explained with a single component susceptibility with hyperfine coefficients and orbital shifts very close to what was predicted by first principles. Thus, the data fully support the previously suggested picture of a susceptibility that appears to be that of a simple metal with a density of states universal to the cuprates in the sense that all strongly overdoped materials have the same density of states and as the doping decreases (in a family specific way) a temperature independent gap opens at the Fermi surface while the states outside the gap remain the same. Due to the temperature dependence of the Fermi function, excitations across the gap lead to the deviations of relaxation and shift (and entropy) from simple metal behavior. It remains to be seen how these findings can be reconciled with the planar Cu data that, e.g., show a very unusual shift anisotropy and a relaxation rate not affected by the pseudogap. 

\section*{acknowledgements}
	We acknowledge the support by the German Science Foundation (DFG HA1893-18-1), and fruitful discussions by other members of our group: Anastasia Aristova, Robin Guehne, Stefan Tsankov. In addition we are thankful to Boris Fine (Leipzig), Andreas Erb (Munich), and Steven Kivelson (Stanford) for their communications on the subject.

\section*{Author contribution}
J.H. introduced the main concepts and had the project leadership; M.A. led the writing and preparation of the manuscript; J.N. led the literature data collection, analyses, and presentation in the
manuscript; all authors contributed to the writing of the manuscript.

\appendix

\section{Data Sources}

\begin{table}[h!]
	\caption{Reference to literature accounts of data. Compounds shown in figures, with critical temperature $T_c$, label and reference link.}
	\begin{tabular}{llll}
		
		Compound&$T_c$&Label&Ref.\\ 
		\hline 
		Bi$_{2}$Sr$_{2}$CaCu$_{2}$O$_{8+x}$ & \SI{82}{K} & Crocker 2011 & \cite{Crocker2011}\\
		YBa$_{2}$Cu$_{3}$O$_{6.6}$ & \SI{60}{K} & Yoshinari 1990 & \cite{Yoshinari1990}\\
		YBa$_{2}$Cu$_{3}$O$_{6.63}$ & \SI{62}{K} & Takigawa 1991, & \cite{Takigawa1991}\\
		 & & Martindale 1998 & \cite{Martindale1998}\\
		YBa$_{2}$Cu$_{3}$O$_{6.8}$ & \SI{84}{K} & Yoshinari 1990 & \cite{Yoshinari1990}\\
		YBa$_{2}$Cu$_{3}$O$_{6.96}$ & \SI{92}{K} & Yoshinari 1990 & \cite{Yoshinari1990}\\
		& & Martindale 1998 & \cite{Martindale1998}\\
		YBa$_{1.92}$Sr$_{0.08}$Cu$_{3}$O$_{7}$& \SI{89}{K} & Horvatic 1993 & \cite{Horvatic1993}\\ 
		YBa$_{2}$Cu$_{4}$O$_{8}$&\SI{81}{K}&Suter 1997&\cite{Suter1997}\\
		La$_{1.76}$Sr$_{0.24}$CuO$_{4}$ & \SI{25}{K} & Zheng 1993 & \cite{Zheng1993}\\
		Tl$_{2}$Ba$_{2}$CuO$_{x}$ & \SI{85}{K} & Kambe 1993 & \cite{Kambe1993}\\
		Tl$_{2}$Ba$_{2}$Ca$_{2}$Cu$_{3}$O$_{10}$ & \SI{125}{K} & Zheng 1996 & \cite{Zheng1996}\\
		\hline 
	
	\end{tabular}
\end{table}


\bibliography{JH-Cuprate.bib}
\printindex
\end{document}